\documentclass[a4paper, aps, amsmath, 10pt, nofootinbib,superscriptaddress]{revtex4-2}
\usepackage[left=3cm,right=3cm,top=3cm,bottom=3cm]{geometry}

\usepackage{graphicx}
\usepackage{booktabs}
\usepackage{tikz}
\usetikzlibrary{arrows.meta}
\usetikzlibrary{shapes.misc}
\tikzset{cross/.style={cross out, draw=black, fill=none, minimum size=2*(#1-\pgflinewidth), inner sep=0pt, outer sep=0pt}, cross/.default={2pt}}
\usepackage{tikz-3dplot}
\usepackage{pgfplots}
\pgfplotsset{compat=1.14}
\usepackage{anyfontsize}
\usepackage{xcolor}
\usepackage{siunitx}
\usepackage{physics}
\usepackage{amsfonts}
\usepackage{mathtools}
\usepackage{svg}
\usepackage{braket}
\usepackage{hyperref}
\usepackage[skip=2pt,font=normalsize]{subcaption}
\usepackage{array}
\usepackage{commath}

\makeatletter
\DeclareRobustCommand{\cev}[1]{%
	{\mathpalette\do@cev{#1}}%
}
\newcommand{\do@cev}[2]{%
	\vbox{\offinterlineskip
		\sbox\z@{$\m@th#1 x$}%
		\ialign{##\cr
			\hidewidth\reflectbox{$\m@th#1\vec{}\mkern4mu$}\hidewidth\cr
			\noalign{\kern-\ht\z@}
			$\m@th#1#2$\cr
		}%
	}%
}
\makeatother

\usepackage[english]{babel}
\begin{document}

\title{Dynamical edge modes in Maxwell theory from a BRST perspective,\\with an application to the Casimir energy}

\author{Fabrizio Canfora}
\email{fabrizio.canfora@uss.cl}
\affiliation{Universidad San Sebastián, sede Valdivia, General Lagos 1163, Valdivia 5110693, Chile}
\affiliation{Centro de Estudios Científicos (CECS), Casilla 1469, Valdivia, Chile}

\author{David Dudal}
\email{david.dudal@kuleuven.be}
\affiliation{KU Leuven Campus Kortrijk -- Kulak, Department of Physics, Etienne Sabbelaan 53 bus 7657, 8500 Kortrijk, Belgium}

\author{Thomas Oosthuyse}
\email{thomas.oosthuyse@kuleuven.be}
\affiliation{KU Leuven Campus Kortrijk -- Kulak, Department of Physics, Etienne Sabbelaan 53 bus 7657, 8500 Kortrijk, Belgium}

\author{Luigi Rosa}
\email{rosa@na.infn.it}
\affiliation{Dipartimento di Matematica e Applicazioni ``R.~Caccioppoli'', Universitá di Napoli Federico II, Complesso Universitario di Monte S.~Angelo, Via Cintia Edificio 5A, 80126 Naples, Italia}
\affiliation{INFN, Sezione di Napoli, Complesso Universitario di Monte S.~Angelo, Via Cintia Edificio 6, 80126 Naples, Italia}

\author{Sebbe Stouten}
\email{sebbe.stouten@kuleuven.be}
\thanks{corresponding author}
\affiliation{KU Leuven Campus Kortrijk -- Kulak, Department of Physics, Etienne Sabbelaan 53 bus 7657, 8500 Kortrijk, Belgium}

\begin{abstract}
Recently, dynamical edge modes (DEM) in Maxwell theory have been constructed using a specific local boundary condition on the horizon. We discuss how to enforce this boundary condition on an infinite parallel plate in the QED vacuum by introducing Lagrange multiplier fields into the action. We carefully introduce appropriate boundary ghosts to maintain BRST invariance. Explicit correspondence of this BRST extended theory with the original DEM formulation is discussed, both directly, and through the correspondence between edge modes and Wilson lines attached to the boundary surface. We then use functional methods to calculate the Casimir energy for the first time with DEM boundary conditions imposed on two infinite parallel plates, both in generalized Coulomb and linear covariant gauge. Depending on the gauge, different fields are contributing, but, after correctly implementing the BRST symmetry, we retrieve the exact same Casimir energy as for two perfectly conducting parallel plates.
\end{abstract}

\maketitle

\section{Motivation}

In this paper, we will study the Casimir effect in the presence of the dynamical edge mode (DEM) boundary conditions recently introduced in \cite{Ball:2024hqe}. 

The Casimir effect \cite{Casimir:1948dh} is a well-known quantum phenomenon revealing the zero-point energy on macroscopic scales. It has been experimentally verified for the first time in \cite{Sparnaay:1958wg}, and has been measured with greater precision in later years \cite{Bordag:2001qi,Lambrecht:1999vd}. The effect also plays an important conceptual role in fields like cosmology and quantum gravity because of its potential influence on spacetime curvature and gravitational interactions \cite{Brevik:2000zb,Brevik:2010okp,vandeKamp:2020rqh} or brane-world models \cite{Saharian:2003qs,Frank:2007jb}. Its applications extend to material science \cite{Jiang:2018ivv,Fukushima:2019sjn,Grushin:2012mt} and nanotechnology \cite{Grushin:2010qoi,Tajik:2022kka,Lopez:2022bky}, where Casimir forces can become significant. 

Edge modes are localized excitations that appear at the boundaries or interfaces of a physical system. 
Famous examples arise in various condensed matter systems: e.g.~in topological insulators \cite{Qi:2010qag,Fidkowski:2010jmn,Zhong2024}, quantum Hall systems \cite{PhysRevB.38.9375,Hatsugai:1993ywa,Cano:2013ooa,Tong:2016kpv}, and Weyl semimetals \cite{Hashimoto,Howard2019,Howard2021}.
However, recently, the term has been used in a slightly different sense in the context of (classical) field theories on manifolds with boundary, more specifically in the study of entanglement entropy \cite{Donnelly:2014fua,Donnelly:2015hxa, Donnelly:2016auv,Donnelly:2020teo}. 
This broader notion of edge mode is of interest to quantum gravity \cite{Huang:2014pfa,Freidel:2020xyx,Ciambelli:2022cfr,Mertens:2022ujr,Donnelly:2022kfs}, high-energy physics \cite{Blommaert:2018oue,Gomes:2018dxs,Gomes:2019xto,Geiller:2019bti,Ball:2024gti}, and lattice field theory \cite{Donnelly:2011hn,Pretko:2015zva,Aoki:2015bsa,Radicevic:2015sza,Soni:2015yga,Chernodub:2023dok}.
In this context, edge modes are defined as extra degrees of freedom living on the entangling surface, which must be introduced in order to obtain a gauge-invariant phase space for the theory on a manifold with boundary. 
Their physical importance lies in the fact that they give a statistical interpretation for the contact term in the entanglement entropy \cite{KABAT1995281,FURSAEV1997697,Solodukhin}. 
The standard manner to introduce these edge modes is to (implicitly) break gauge invariance at the boundary, which ``releases'' previously gauge and thus unphysical degrees of freedom into extra physical degrees of freedom on the boundary. 
It should be noted that it is not clear exactly how the notion of edge modes in the context of entanglement entropy relates to the more conventional edge modes from condensed matter physics, see also \cite[Sect.~5.9]{Riello:2020zbk}. Henceforth, we will be using the term ``edge mode'' in the former sense of would-be gauge boundary degrees of freedom on an entangling surface.

In Yang-Mills (or Maxwell) theory, the edge modes are associated with a choice of gauge on the boundary, or thus with the normal electrical flux through the boundary.\footnote{A different line of research instead focuses on gauge-covariant super-selection sectors of the electrical flux through the boundary. An advantage of this approach would be that no extra degrees of freedom are needed and that the procedure does not depend on the chosen gauge on the boundary, making it more canonical. See e.g.~\cite{Gomes:2019xto,Riello:2020zbk,Riello:2021lfl}.}
In the seminal papers \cite{Donnelly:2014fua,Donnelly:2015hxa,Donnelly:2016auv}, these edge modes were added to the path integral by hand, but in \cite{Ball:2024hqe} it was shown that they arise naturally when one imposes some specific boundary conditions (DEM conditions).\footnote{The ``dynamical'' in DEM, as coined in \cite{Ball:2024hqe}, refers to the mere observation that the edge modes are a dynamical consequence of the DEM boundary conditions an sich. Throughout our paper, we will come back to the dynamics, in terms of their time dependence, of the edge modes themselves.}
For completeness, we mention some other work regarding non-trivial boundary conditions and edge modes: \cite{David:2021wrw,Mukherjee:2023ihb,Cheng:2023bcv,Cheng:2023cms,Ball:2024xhf}.

In the present paper, we will study the Casimir effect for a system of two infinite parallel plates on which we impose this set of boundary conditions.

Our methodology is as follows. 
We start by incorporating the DEM boundary conditions into the action using Lagrange multiplier fields
\cite{Dudal:2020yah,Canfora:2022xcx,Oosthuyse:2023mbs,Dudal:2024PEMC,Dudal:2024Robin}, see also \cite{Bordag:1983zk}. Since DEM conditions are not compatible with the full set of gauge transformations with non-trivial boundary support, appropriate boundary ghost fields, to maintain explicit BRST invariance (see \cite{Becchi:1974md,Becchi:1974xu,Becchi:1975nq,Tyutin:1975qk}), are needed. Excellent pedagogical reviews on BRST symmetry can be found in \cite{Baulieu:1983tg,Henneaux:1994lbw,Piguet:1995er,Fuster:2005eg}.
Finally, using path integral techniques, we retrieve the Casimir energy by integrating out the fields one-by-one.

The paper is organized as follows. In section \ref{sec:action}, we define DEM boundary conditions and the action of interest; in section \ref{sec:BRST}, we note down a BRST invariant extension of this action; and in section \ref{sec:correspondence}, we validate that the BRST invariant extension can still be interpreted as describing the dynamics of edge modes as in \cite{Ball:2024hqe}. Next, we calculate the Casimir energy for the DEM parallel plate setup in generalized Coulomb gauge in section \ref{sec:casimirCoulomb1}, and in linear covariant gauge in section \ref{sec:casimirLandau1}. We end by discussing our conclusions and giving an outlook in section \ref{ch:conclusion}.

\section{Setup}\label{ch:setup}
\subsection{Action and boundary conditions}\label{sec:action}
We will consider flat spacetime in 3+1D, but for notational simplicity, we will perform a Wick-rotation from the start, leaving us with 4D Euclidean space. Let us use the convention that Greek indices \(\mu,\nu,...\) run over all 4 dimensions \(\{t,x,y,z\}\), whereas Roman indices \(i,j,...\) only run over \(\{x,y,z\}\), and Roman indices \(a,b,...\) run over \(\{x,y\}\). In what follows, we will always write Euclidean indices as lower indices, understanding a summation over repeated indices.

The theory of interest is Maxwell theory, given by the Euclidean action
\begin{equation*}
	S_\text{Maxwell} = \int \dd[4]{x} \frac{1}{4} F_{\mu\nu} F_{\mu\nu},
\end{equation*}
where the Maxwell field strength is denoted by \(F_{\mu\nu} = \partial_\mu A_\nu - \partial_\nu A_\mu\).

Recall that the Maxwell action exhibits the gauge freedom \(A_\mu \rightarrow A_\mu + \partial_\mu \chi\). In this paper, we will consider two possible classes of gauge fixing conditions: linear covariant gauge and generalized Coulomb gauge. These are implemented into the action by
\begin{equation*}
	S_\text{GF} = \int \dd[4]{x} \left( h \mathcal{F}[A] + \frac{\xi}{2} h^2 \right),
\end{equation*}
in which \(\xi\) represents the gauge parameter, and \(h\) the Nakanishi-Lautrup field \cite{Nakanishi:1966zz,Lautrup1967canonical,Nakanishi:1990qm}. The gauge fixing functional is represented by \(\mathcal{F}[A]\), meaning that for the linear covariant gauges \(\mathcal{F}[A] = \partial_\mu A_\mu\) and for Coulomb gauges \(\mathcal{F}[A] = \partial_i A_i\). The choice $\xi=0$ corresponds to, respectively, the standard Landau (aka.~Lorenz) and Coulomb gauge.

\begin{figure}
	\centering
	\includegraphics{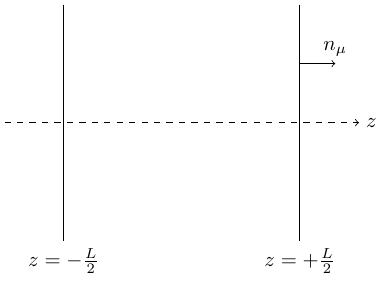}
	\caption{Schematic representation of the boundary configuration: infinitely large plates at \(z=\pm \frac{L}{2}\) with normal vector \(n_\mu = (0,0,0,1)\).}
	\label{fig:boundary}
\end{figure}

Now, we place two parallel plates in this flat Euclidean space, both infinitely large and infinitely thin. They are located at \(z=z^-\equiv-\frac{L}{2}\) and \(z=z^+\equiv+\frac{L}{2}\), for some \(L>0\) (see Fig.~\ref{fig:boundary}). Denote the normal vector on the plates by \(n_\mu = (0,0,0,1)\). Note that, technically speaking, we are not considering a bounded manifold: the plates do not correspond to the end of space, but rather they represent interfaces or defects placed in unbounded 4D Euclidean space.\footnote{During recent years, the study of interfaces/defects in quantum field theories has seen increased interest, mostly in the context of conformal field theories and/or holography \cite{Andrei:2018die}.}

Let us briefly recall that perfect electric conductor (PEC) and perfect magnetic conductor (PMC) \cite{Edery:2008rj,Edery:2009vr} boundary conditions are given by
\begin{align*}
	\text{PEC}: \quad &\widetilde{F}_{\mu\nu}n_\nu = 0\\
	\text{PMC}: \quad &F_{\mu\nu}n_\nu = 0
\end{align*}
where the dual field strength is defined as
\begin{equation*}
	\widetilde{F}_{\mu\nu} = \frac{i}{2} \varepsilon_{\mu\nu\alpha\beta} F_{\alpha\beta},
\end{equation*}
in which the factor \(i\) accounts for the Wick-rotation of the Levi-Civita tensor \(\varepsilon\). PEC and PMC conditions are also known as the relative and the absolute boundary condition respectively \cite{Vassilevich:2003xt}, and in this context the PEC boundary conditions are often expressed as \(A_\mu = 0\).\footnote{Making explicit the connection between these two versions of PEC necessitates boundary conditions imposed on the gauge rotation angles as well, see \cite{Vassilevich:2003xt}.} Following \cite{Ball:2024hqe}, we can now define DEM conditions by taking the PEC condition for the \(t\)-direction and PMC conditions for the other directions:
\begin{equation*}
	\text{DEM}: \quad
	\begin{cases}
		A_t = 0 \\
		F_{i\nu}n_\nu = 0
	\end{cases}.
\end{equation*}
We will enforce these DEM conditions on both plates by introducing Lagrange multiplier fields \(b^\pm_t\) and \(b^\pm_a\). The boundary term in the action thus becomes
\begin{equation*}
	S_\text{BC} = \int\dd[4]{x} \Big( b^\rho_t(\cev{x}) A_t + b^\rho_a(\cev{x}) F_{az} \Big) \delta(z-z^\rho) ,
\end{equation*}
in which we have introduced upper indices \(\rho,\sigma,...\) that run over the 2 plates \(\{-,+\}\). For a given four-vector \(v\), we will use the notation \(\vec{v} = (v_x,v_y,v_z)\) and \(\cev{v} = (v_t,v_x,v_y)\). We remind the reader that indices \(a,b,...\) only run over \(\{x,y\}\).

At this point, we should note that the DEM conditions break the full Maxwell gauge invariance on the boundary: the local gauge transformation \(A_\mu \rightarrow A_\mu + \partial_\mu \chi\) only conserves the DEM conditions if 
\begin{equation}\label{extravoorwaarde}
    \partial_t \chi|_{z=\pm \frac{L}{2}} = 0,
\end{equation} 
but otherwise ``large gauge transformations''\footnote{In the edge mode literature, ``large gauge transformations'' or ``asymptotic gauge transformations'' refer to those with non-vanishing support on the boundary.} are allowed. These will give rise to an extra scalar degree of freedom, the edge mode, living on the boundary.

We will introduce this new scalar field in section \ref{sec:BRST} using a BRST invariance argument. If we want physical quantities to be gauge invariant, we need to make sure that the action itself is gauge invariant. In fact, as is discussed in \cite{Moss:1996ip,Vassilevich:1997iz,Vassilevich:2003xt,Acharyya:2016xaq}, to quantize gauge theories in manifolds with spatial boundaries, one needs to include the ghost fields in the discussion. As such, the requirement of gauge invariance must be replaced by BRST invariance. This will be discussed in the next section. For completeness, let us mention that in \cite{Baulieu:2024oql,Baulieu:2024rfp}, further aspects of BRST invariance were discussed for manifolds with boundaries, in particular in relation to Noether's theorem.

\subsection{BRST invariance}\label{sec:BRST}
At this moment, the action consists of three terms
\begin{equation}\label{eq:action-no-ghosts}
	S_\text{Maxwell}+S_\text{GF}+S_\text{BC}=\int \dd[4]{x} \left[ \frac{1}{4} F_{\mu\nu} F_{\mu\nu} + \left( h \mathcal{F}[A] + \frac{\xi}{2} h^2 \right) + \Big( b^\rho_t A_t + b^\rho_a F_{az} \Big) \delta(z-z^\rho) \right].
\end{equation}
We can introduce the ghost fields \(c\) and \(\bar{c}\) in the standard manner by adding the ghost term
\begin{equation}\label{eq:action-ghost}
	S_\text{ghost} = \int \dd[4]{x} \bar{c} \frac{\delta \mathcal{F}[A]}{\delta A_\mu} \partial_\mu c.
\end{equation}
Let us list the BRST transformation \(s\) for all of our fields, gathering doublets column per column:
\begin{align*}
	sA_\mu&=-\partial_\mu c & s\bar{c}&=h & sb^\rho_t&=0 & sb^\rho_a&=0 \\
	sc&=0 & sh&=0 & & & &
\end{align*}
It clearly is nilpotent, \(s^2=0\). Inspecting the BRST transformation of our action term-by-term, we can easily see that \(s\left( \frac{1}{4} F_{\mu\nu} F_{\mu\nu} \right) = 0\), and that \(s\left( b^\rho_a F_{az} \right) = 0\). Moreover, we can write
\[h \mathcal{F}[A] + \frac{\xi}{2} h^2 + \bar{c} \frac{\delta \mathcal{F}[A]}{\delta A_\mu} \partial_\mu c = s \left( \bar{c} \frac{\delta \mathcal{F}[A]}{\delta A_\mu} A_\mu + \frac{\xi}{2} \bar{c} h \right),\]
a BRST exact term, such that \(s\left(h \mathcal{F}[A] + \frac{\xi}{2} h^2 + \bar{c} \frac{\delta \mathcal{F}[A]}{\delta A_\mu} \partial_\mu c \right) = 0\) as well. That only leaves \(s\left( b^\rho_t A_t\right) = -b^\rho_t \partial_t c \neq 0\). Inspired by \cite{Vassilevich:2003xt,Baulieu:2024oql}, we seek to cancel this term by introducing extra multiplier fields to enforce Dirichlet boundary conditions on the ghosts \((c,\bar{c})\) and \(h\): a pair of Grassmann ghost fields \((\eta,\bar{\eta})\) and a scalar field \(\gamma\). This results in the final term we will add to the action
\begin{equation}\label{ghostaction}
	S_\text{ghBC} = \int \dd[4]{x} \left( \bar{\eta}^\rho(\cev{x}) c + \eta^\rho(\cev{x}) \bar{c} + \gamma^\rho(\cev{x}) h \right) \delta(z-z^\rho).
\end{equation}
These extra boundary constraints on ghosts and multipliers also entered the recent work \cite{Baulieu:2024rfp} to ensure well-defined ``large'' charges and their algebra.

Let us write down the complete BRST transformation including these new fields
\begin{align}\label{fullBRST }
	sA_\mu&=-\partial_\mu c & s\bar{c}&=h & s\gamma^\rho&=\eta^\rho & s\bar{\eta}^\rho&=-\partial_t b^\rho_t & sb^\rho_a&=0 \\
	sc&=0 & sh&=0 & s\eta^\rho&= 0 &sb^\rho_t&=0 & & & & \nonumber
\end{align}
 Clearly, $s^2=0$, and the full action is BRST invariant, even off-shell. Indeed, one can easily check that \(\eta^\rho \bar{c} + \gamma^\rho h = s\left( \gamma^\rho \bar{c} \right)\), and that \(s\left( b^\rho_t A_t + \bar{\eta}^\rho c \right) = 0\). Important to note is that this last expression \(b^\rho_t A_t + \bar{\eta}^\rho c\) is BRST closed, but not BRST exact, meaning it cannot be written as \(s(...)\). This means that it is not a pure gauge fixing term and thus describes some real physical content, that is, the edge mode sector.

It is beyond the scope of the current paper, but it would be instructive to construct the Fock space by canonical quantization and projecting to the physical subspace, i.e.~the non-trivial part of the BRST  cohomology with ghost number zero. From the functional expressions \eqref{fullBRST }, many degrees of freedom are BRST trivial as being doublet partners \cite{Piguet:1995er} or will cancel under the quartet mechanism \cite{Kugo:1979gm}. The physical subspace seems to consist of the usual two transverse gauge polarizations, supplemented with the extra degree of freedom encoded in the $b_t$-sector, at least if $\partial_t b_t= 0$. The latter is clearly reminiscent of the admissibility condition \eqref{extravoorwaarde}.

For the record, based on \eqref{extravoorwaarde}, one might be tempted to replace \eqref{ghostaction} with
\begin{equation*}
	S_\text{ghBC}^{\text{wrong}} = \int \dd[4]{x} \left( \bar{\eta}^\rho(\cev{x}) \partial_t c + \eta^\rho(\cev{x}) \partial_t\bar{c} + \gamma^\rho(\cev{x}) \partial_t h \right) \delta(z-z^\rho).
\end{equation*}
which, together with \eqref{eq:action-no-ghosts} and \eqref{eq:action-ghost}, is also BRST invariant under \eqref{fullBRST }  up to the replacement
\begin{align}\label{fullBRST 2}
	 s\bar{\eta}^\rho&= b^\rho_t & sb^\rho_t&=0.
\end{align}
However, this makes no sense. Indeed, this would imply that the would-be edge mode $b^\rho_t$ itself then becomes part of a BRST doublet and thence physically trivial. In fact, note that in such case $\int \dd[4]{x} \left( b^\rho_t A_t+\bar{\eta}^\rho \partial_t c\right)\delta(z-z^\rho)=s\int \dd[4]{x} \left(\bar{\eta}^\rho A_t\right)\delta(z-z^\rho)$, which corresponds to nothing more than the temporal (Weyl) gauge on the boundary, with $b^\rho_t$ playing the role of the Nakanishi-Lautrup multiplier.

For completeness and later usage, we write out the full (correct) BRST invariant action once more:
\begin{align}
	\begin{split}\label{eq:action}
	S &= S_\text{Maxwell}+S_\text{GF}+S_\text{BC}+S_\text{ghost}+S_\text{ghBC} \\
	&= \int \dd[4]{x} \bigg[ \frac{1}{4} F_{\mu\nu} F_{\mu\nu} + \left( h \mathcal{F}[A] + \frac{\xi}{2} h^2 \right) + \Big( b^\rho_t A_t + b^\rho_a F_{az} \Big) \delta(z-z^\rho) \\
	&\qquad \qquad \qquad + \bar{c} \frac{\delta \mathcal{F}[A]}{\delta A_\mu} \partial_\mu c + \left( \bar{\eta}^\rho c + \eta^\rho \bar{c} + \gamma^\rho h \right) \delta(z-z^\rho) \bigg]
	\end{split}
\end{align}

\subsection{Correspondence to dynamical edge modes}\label{sec:correspondence}
Having found a BRST invariant action that enforces the DEM conditions on the parallel plates, we need to make sure that this action is still describing the dynamical edge modes discussed in \cite{Ball:2024hqe}. We will make this correspondence explicit in two steps: we first show (in two ways) that in our formalism, the edge field is embodied by the multiplier field \(b_t\). Secondly, we will show that our action \eqref{eq:action} describes edge mode dynamics relatable to the dynamics found in \cite{Ball:2024hqe}.

\subsubsection{Correspondence of edge fields -- Gauss's law}
In \cite{Ball:2024hqe}, the edge field is given by the boundary field \(E_\perp\), which corresponds to the \(z\)-direction for our plate configuration. So our goal is to show that, on-shell, \(E_\perp\) corresponds to \(b_t\). In order to make this correspondence explicit, we need to inspect the equations of motion for our action \eqref{eq:action}. Let us start with the equation of motion for \(h\):
\begin{equation}\label{eq:div-A}
	\frac{\delta S}{\delta h} = 0 \Leftrightarrow \mathcal{F}[A] + \gamma^\rho \delta(z-z^\rho) = 0.
\end{equation}
Using this identity, the equation of motion for \(A_t\) simplifies to
\begin{equation}\label{eq:EOMAt}
	\frac{\delta S}{\delta A_t}=0 \Leftrightarrow  \partial^2 A_t - \partial_t \partial_\mu A_\mu = b_t^\rho(\cev{x}) \delta(z-z^\rho).
\end{equation}
Now we can write down Gauss's law (remembering that \(E_i = \partial_i A_t - \partial_t A_i\))
\begin{align}
	\partial_i E_i &= \nabla^2 A_t - \partial_t \partial_i A_i \nonumber\\
	&= \partial^2 A_t - \partial_t \partial_\mu A_\mu \nonumber\\
	&= b_t^\rho(\cev{x}) \delta(z-z^\rho), \label{eq:Gauss}
\end{align}
where we have used \eqref{eq:EOMAt} in the last identity. Note that the right hand side is a pure boundary term. We have thus found that the multiplier field \(b_t\) plays the role of an electrical flux source on the boundary. In fact, \(b_t\) is nothing else than the edge mode.

\begin{figure} 
	\centering
	\includegraphics{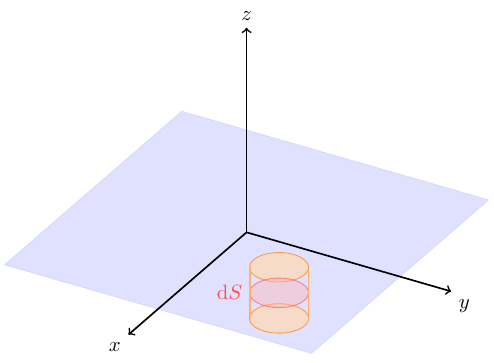}
	\caption{An infinitesimal cylinder \(\Sigma\) bounding a volume \(V\) piercing one of the plates.}
	\label{fig:cylinder}
\end{figure}

We can make this even more clear by using a standard pillbox reasoning. Since \(\partial_i E_i\) is only non-zero on the plates, let's consider Gauss's divergence theorem on a small cylinder \(\Sigma\) piercing one of the plates (say at \(z=z^\sigma\)), see Fig.~\ref{fig:cylinder}. Denote the volume enclosed by \(\Sigma\) with \(V\). On the one hand, we thus have
\begin{equation}\label{shorthand}
	\int_V \dd[3]{V} \partial_i E_i =  \int_{\dd S}\dd[2]{\!A} n_i E_i = (E_z^+-E_z^-) \dd S\equiv E_\perp \dd S,
\end{equation}
where we have used that the surface \(\dd S\) is infinitesimal. The $\pm$-superscripts correspond to the two sides of the boundary.

On the other hand, we can use \eqref{eq:Gauss}:
\begin{equation*}
	\int_V \dd[3]{V} \partial_i E_i = \int_V \dd[3]{V} b_t^\rho \delta(z-z^\rho) = \int_{\dd S} \dd[2]{\!A} b_t^\sigma = b_t^\sigma \dd S.
\end{equation*}
Comparing both expressions, we find
\begin{equation*}
	b_t^\sigma =  E_\perp\qquad\textrm{on the plate at \(z=z^\sigma\)}.
\end{equation*}
This makes the correspondence of the edge modes \(E_\perp\) and \(b_t\) quite explicit. To our understanding, this interpretation of the edge field \(b_t\) as a boundary source of electric flux corresponds to the language of \cite{Riello:2020zbk,Riello:2021lfl}. In fact,  \eqref{eq:Gauss} leads to
\begin{equation}\label{eq:Gaussdt}
	\partial_t\partial_i E_i = \partial_t b_t^\rho(\cev{x}) \delta(z-z^\rho) = s(\bar{\eta}^\rho(\cev{x}) \delta(z-z^\rho)), 
\end{equation}
showing that the physical part of the flux is non-dynamical, as also discussed in \cite{Riello:2021lfl}. Indeed, at the level of physical states, the BRST exact r.h.s.~of \eqref{eq:Gaussdt} mods out to zero.

Let us also discuss the canonically conjugate field to the edge mode. In \cite{Ball:2024hqe}, the photon field in Coulomb gauge is parametrized as
\begin{equation}\label{eq:parametrization-A}
	A_i = \tilde{A}_i + \partial_i \alpha
\end{equation}
where \(\tilde{A}\) has the properties \(\partial_i \tilde{A}_i = 0\) and \(n_i \tilde{A}_i|_{z=\pm\frac{L}{2}}=0\). It is shown that this \(\alpha\) is the canonically conjugate field to the edge mode \(E_\perp\). We thus wish to find the field that \(\alpha\) corresponds to in our formalism. For the sake of simplicity, let us show the correspondence for a configuration of only one plate. The generalization to a two-plate configuration can be made, but the expressions are less clean because the edge modes of both plates mix up.

Taking the divergence of \eqref{eq:parametrization-A} gives us that \(\partial_i A_i = \vec{\partial}^2\alpha\), or, by virtue of \eqref{eq:div-A} in Coulomb gauge,
\begin{equation}\label{vglalpha}
	-\gamma \delta(z-z_\text{plate}) = \vec{\partial}^2\alpha.
\end{equation}
By integrating this equation along the normal direction through the plate, this automatically gives for the solution that it will be subject to
\begin{equation}\label{neumann}
	-\gamma = \partial_z \alpha^+-\partial_z \alpha^- \equiv \partial_\perp \alpha
\end{equation}
on the plate, where we introduced a shorthand similar to \eqref{shorthand}. This identity also follows from the decomposition \eqref{eq:parametrization-A}, giving $\left. A_\perp\right\vert_\text{plate}\equiv\left.A_z^+-A_z^-\right\vert_\text{plate}= \left. \partial_\perp\alpha\right\vert_\text{plate}$, whilst another pillbox reasoning applied to \eqref{eq:div-A} boils down to $\left.A_\perp\right\vert_\text{plate}=-\gamma$. Note that \eqref{neumann}, or thus $\left.A_\perp\right\vert_\text{plate}=\left. \partial_\perp\alpha\right\vert_\text{plate}$, corresponds to the imposed end-of-space Neumann boundary data of \cite{Ball:2024hqe}, something which arises again quite naturally in our BRST invariant approach. 

Using the Green's function for the 3D Laplacian \(\frac{-1}{4\pi |\vec{x}|}\), we can invert this relation \eqref{vglalpha} explicitly as
\begin{align*}
	\alpha(x) &= -\frac{1}{\vec{\partial}^2} \left( \gamma(\cev{x}) \delta(z-z_\text{plate}) \right) \\
	&\propto \int \dd{x'} \dd{y'} \frac{\gamma(t,x',y')}{\sqrt{(x-x')^2+(y-y')^2+(z-z_\text{plate})^2}}.
\end{align*}
Given that the quantum field $\gamma$ is supposed to fall off sufficiently fast at infinity, the foregoing integral is well-behaved and we can assume that the field $\alpha$ also drops to zero at infinity, which tacitly sets the boundary condition for \eqref{vglalpha}. 

On the plate, we find
\begin{align}
	\alpha(\cev{x},z_\text{plate}) &\propto \int \dd{x'} \dd{y'} \frac{\gamma(t,x',y')}{\sqrt{(x-x')^2+(y-y')^2}} \nonumber\\
	&\propto \frac{1}{\sqrt{-\nabla^2_{2D}}} \gamma(\cev{x}). \label{eq:alpha-gamma-correspondence}
\end{align}
The last identification is most easily appreciated from the fact that the 2D Fourier transform of \(1/\sqrt{(x-x')^2+(y-y')^2}\) is proportional to \(1/\sqrt{k_x^2+k_y^2}\), that is, indeed corresponding to \(1/\sqrt{-\nabla^2_{2D}}\). We can thus conclude that in our formalism, \(\gamma\) takes up the role of the conjugate edge mode \(\alpha\). It is satisfying to see that demanding BRST invariance of the action \eqref{eq:action-no-ghosts} led us to introducing the ``missing'' conjugate edge field.

\subsubsection{Correspondence of edge fields -- Wilson lines ending on the boundary}
In this paragraph, we will give another argument why \(b_t\) should be interpreted as the edge mode, this time using Wilson lines ending on the boundary, in the spirit of e.g.~\cite{Donnelly:2016auv,Blommaert:2018oue,Geiller:2019bti,Riello:2020zbk,Riello:2021lfl,Carrozza:2021gju,Kabel:2023jve}. Let us again consider a configuration of only one plate for simplicity. Imagine that at \(t=t_0\) we want to place two opposite point charges \(\pm q\) on the plate at \((x_\pm,y_\pm,z_\text{plate})\). The source \(j\) corresponding to this configuration is given by
\begin{equation*}
	j_\mu = q \Big( \delta(x-x_+) \delta(y-y_+) - \delta(x-x_-) \delta(y-y_-) \Big) \delta(z-z_\text{plate})\theta(t-t_0)\delta_{\mu t},
\end{equation*}
where \(\theta\) is the Heaviside step function. As such, the term that needs to be added to the action is
\begin{equation}\label{eq:action-point-charges}
	\int \dd[4]{x} A_\mu j_\mu = q\int_{t_0}^\infty \dd{t} \bigg[ A_t(x_+,y_+,z_\text{plate}) - A_t(x_-,y_-,z_\text{plate}) \bigg].
\end{equation}
We will discuss two possible ways to include such a term into the action.

\begin{figure}
	\centering
	\includegraphics{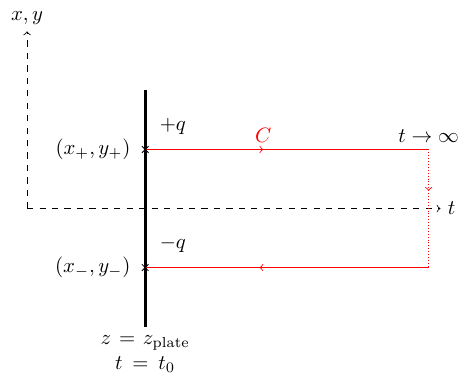}
	\caption{Schematic representation of the Wilson line \(W_C\): starting from \((x_+,y_+,z_\text{plate})\) on the plate at \(t=t_0\), letting \(t\rightarrow\infty\), then going to the point \((x_-,y_-,z_\text{plate})\) on the plate, and finally returning to \(t=t_0\).}
	\label{fig:Wilson-line}
\end{figure}

On the one hand, we can introduce a temporally stretched Wilson line ending on the plate. More precisely, consider the piece-wise straight curve \(C\) connecting \[(t_0,x_+,y_+,z_\text{plate}) \rightarrow (t=\infty,x_+,y_+,z_\text{plate}) \rightarrow (t=\infty,x_-,y_-,z_\text{plate}) \rightarrow (t_0,x_-,y_-,z_\text{plate}),\] as depicted in Fig.~\ref{fig:Wilson-line}. In general, the Wilson line \(W_C\) following \(C\) is defined as
\begin{equation*}
	W_C = \exp\left(q\int_C \dd[4]{x_\mu} A_\mu\right).
\end{equation*}
Despite not being a loop, the Wilson line with endpoints on the boundary is also BRST invariant since \(c|_\text{plate} = 0 \):
\begin{equation*}
	s \left(W_C\right) =  W_C \cdot q\int_C \dd[4]{x_\mu} \partial_\mu c = 0.
\end{equation*}
Following \cite{Fischler:1977yf,Dudal:2009ti}, we note that at temporal infinity \(F_{\mu\nu}^2\) must vanish to ensure finite action, such that the photon field \(A_\mu\) is pure gauge, and we can set \(A_\mu=0\) at \(t=\infty\). This means that the Wilson line \(W_C\) only contains two non-vanishing pieces, yielding the sought-after contribution \eqref{eq:action-point-charges} to the action:
\begin{equation*}
	 W_C =  \exp\left(q\int_{t_0}^\infty \dd{t} \bigg[ A_t(x_+,y_+,z_\text{plate}) - A_t(x_-,y_-,z_\text{plate}) \bigg]\right).
\end{equation*}

On the other hand, we can plug the classical background \(j_\mu\) into the boundary field \(b_t\). Indeed, if we replace \(b_t \delta(z-z_\text{plate}) \rightarrow b_t \delta(z-z_\text{plate}) + j_t\) in the action \eqref{eq:action}, then we obtain the same sought-after action term \eqref{eq:action-point-charges}. After this substitution, we can interpret \(b_t\) as describing quantum fluctuations around the classical field \(j_\mu\).\footnote{Here again, we find the interpretation of \(b_t\) being a boundary source of electric flux, as in \cite{Riello:2020zbk,Riello:2021lfl}.} As such, the classical limit of the boundary field \(b_t\) corresponds to the end points (classical charges) of a Wilson line ``anchored'' on the plate. We have thus recovered the correspondence between edge modes and (explicitly BRST invariant) Wilson lines attached to the boundary surface, see also \cite{Blommaert:2018rsf} for a discussion of such in a somewhat different context. This shows once more that in our formalism, the edge mode is given by \(b_t\).

\subsubsection{Correspondence of edge dynamics}\label{sec:edgeDynamics}
Now we can investigate the dynamics of the edge modes. More specifically: we want to show that the edge part of our action corresponds to the edge Hamiltonian in \cite[Eq.~(2.43)]{Ball:2024hqe}. Again, for the sake of simplicity, let us show the correspondence for a configuration of only one plate. For the one-plate configuration, the edge Hamiltonian becomes \cite[Eq.~(6)]{Nair:2022oqk}
\begin{equation}\label{eq:edgeHam}
	H_\text{edge} = \int_\text{plate} \dd[3]{\cev{x}} E_\perp \frac{1}{\sqrt{-\nabla^2_{2D}}} E_\perp,
\end{equation}
where \(E_i = \partial_i A_t - \partial_t A_i\) is the electric field, and \(\nabla^2_{2D} = \partial_x^2 + \partial_y^2 \) is the 2D Laplacian.

This expression for the edge Hamiltonian has been derived in Coulomb gauge with gauge parameter \(\xi = 0\). However, in our derivation, we will postpone choosing a gauge for as long as possible.

Inspecting the action \eqref{eq:action}, we see that the ghosts \(c,\bar{c},\eta,\bar{\eta}\) decouple from the other fields. Since we are only interested in the dynamics of the edge mode \(b_t\), we can thus ignore these ghosts terms for the moment. To arrive at a boundary action, we want to integrate out the Nakanishi-Lautrup field \(h\) and the photon field \(A_\mu\). The former can be readily integrated out, after which we get
\begin{equation}\label{eq:action-h-out}
	S_{A,b,\gamma} = \int \dd[4]{x} \bigg[ \frac{1}{4} F_{\mu\nu} F_{\mu\nu} + \frac{1}{2\xi} \Big( \mathcal{F}[A] + \gamma\delta(z-z_\text{plate}) \Big)^2 + \Big( b_t A_t + b_a F_{az} \Big) \delta(z-z_\text{plate}) \bigg].
\end{equation}
Next, we have to integrate out the photon field \(A\). This derivation will depend on the choice of gauge function \(\mathcal{F}\), so from here on we specify to work in generalized Coulomb gauge, i.e.~\(\mathcal{F}[A] = \partial_i A_i\). We will postpone the explicit calculation of this path integral to section \ref{sec:casimirCoulomb1}, where we will return to this issue. At this point, it suffices to state the form of the resulting boundary action:
\begin{equation}\label{eq:2D-lapl}
	S_{b,\gamma} = \int \dd[3]{\cev{x}} \bigg[ b_t \frac{1}{\sqrt{-\nabla^2_{2D}}} b_t + b_a \mathbb{N}_{ab} b_b  + b_t j_t \bigg],
\end{equation}
for some operator \(\mathbb{N}\), and source \(j_t \propto \frac{\partial_t}{\sqrt{-\nabla^2_{2D}}} \gamma\). From this form, it is clear to see that \(b_a\) decouples from the \((b_t,\gamma)\)-sector. After partially integrating the source term and noting that \(\frac{1}{\sqrt{-\nabla^2_{2D}}}\gamma\) is the canonically conjugate field to \(b_t\) (i.e.~\(\frac{\delta \mathcal{L}}{\delta \partial_t b_t} = \frac{1}{\sqrt{-\nabla^2_{2D}}}\gamma\))\footnote{After \eqref{eq:alpha-gamma-correspondence}, this is a second confirmation that \(\frac{1}{\sqrt{-\nabla^2_{2D}}}\gamma\) corresponds to the conjugate edge mode \(\alpha\) in \cite{Ball:2024hqe}.}, the Lagrangian for the edge mode \(b_t\) has been brought in the form
\begin{align*}
	\mathcal{L} &= b_t \frac{1}{\sqrt{-\nabla^2_{2D}}} b_t + \frac{1}{\sqrt{-\nabla^2_{2D}}}\gamma \partial_t b_t \\
	&= \mathcal{H} + \pi \dot{\varphi},	
\end{align*}
where \(\mathcal{H}\) is the edge Hamiltonian density in \eqref{eq:edgeHam}. As such, we have indeed related both formulations of the edge dynamics.

\section{The Casimir energy: two gauges}\label{sec:casimir}
Now that we have established the correspondence between the BRST invariant action \eqref{eq:action} and the edge Hamiltonian \eqref{eq:edgeHam}, we can investigate the Casimir effect for the parallel plate configuration with DEM conditions.
The methodology we will use has been developed in \cite{Dudal:2020yah,Canfora:2022xcx,Oosthuyse:2023mbs,Dudal:2024PEMC,Dudal:2024Robin}.
Put succinctly, the method can be summarized as follows.
Let us denote the (infinite) spacetime integration volume with \(\mathcal{V}=\ell_t\ell_x\ell_y\ell_z\), with $\ell_{t,x,y,z}\to+\infty$.
Then we have the path integral identity
\begin{equation*}
	Z = \int \mathcal{D}\Phi e^{-S[\Phi]} = e^{-\ell_t E}
\end{equation*}
with \(E\) the vacuum energy of the theory, and \(\Phi\) a shorthand for all the fields.
In a fully translationally invariant theory one could then find the energy density \(\mathcal{E}\) from \(E=\ell_x \ell_y \ell_z \mathcal{E}\).
However, care must be taken as we have both 4D ``bulk'' \(\Phi_\text{bulk}=\{A_\mu, h, c, \bar{c}\}\) and 3D ``plate'' \(\Phi_\text{plate}=\{b, \gamma, \eta, \bar{\eta}\}\) fields.
Accordingly, the vacuum energy \(E\) will contain contributions from bulk and plate fields, the latter of which will correspond to the Casimir energy.
Schematically, as our action is quadratic in the fields, with plate fields acting as sources for the bulk fields, it can be diagonalized by a shift of the bulk fields (which leaves the measure invariant \(\mathcal{D}\Phi_\text{bulk}=\mathcal{D}\Phi^\prime_\text{bulk}\)):
\begin{align*}
    Z &= \int \mathcal{D}\Phi_\text{bulk} \mathcal{D}\Phi_\text{plate} e^{-S[\Phi_\text{bulk},\Phi_\text{plate}]} \\
    &= \int \mathcal{D}\Phi^\prime_\text{bulk} e^{-S_\text{bulk}[\Phi^\prime_\text{bulk}]} \int \mathcal{D}\Phi_\text{plate} e^{- S_\text{plate}[\Phi_\text{plate}]} \\
    &= e^{-\ell_t E_\text{bulk}} e^{-\ell_t E_\text{Cas}}
\end{align*}
where the bulk action \(S_\text{bulk}\) does not depend on the plate configuration.
The plate action only contains the plate fields, hence describing a 3D theory, and the Casimir energy density follows as \(E_\text{Cas} = \ell_x \ell_y \mathcal{E}_\text{Cas}\).

The method thus boils down to finding an expression for the plate action starting from the complete action \(S\) in \eqref{eq:action}, and calculating the functional determinant arising from integration over the plate fields to obtain the Casimir energy.
Both the construction of the plate action and the functional determinant require regularization and are worked out in detail in the following section and appendices.

As already mentioned in section \ref{sec:edgeDynamics}, the ghosts decouple from the other fields, meaning that we can investigate both sectors separately: \(Z = Z_{c,\bar{c},\eta,\bar{\eta}} Z_{A,b,h,\gamma}\). For the non-ghost part, we have already integrated out \(h\), yielding the action \eqref{eq:action-h-out}. In order to integrate out the other fields, we will need to work with a specified gauge function \(\mathcal{F}[A]\). We will do so for two common choices: first, the generalized Coulomb gauge \(\mathcal{F}[A] = \partial_i A_i\), in which the analysis in \cite{Ball:2024hqe} was done; second, the linear covariant gauge \(\mathcal{F}[A] = \partial_\mu A_\mu\). Of course, both choices yield the same result, per BRST  invariance, but it is interesting to see that different fields contribute differently to the vacuum energy depending on the gauge.

We will discuss Coulomb gauge in section \ref{sec:casimirCoulomb1} and linear covariant gauge in section \ref{sec:casimirLandau1}. For the reader's convenience, we will only present the most relevant formulae and results below, relegating the details to the Appendix.

\subsection{Casimir energy in generalized Coulomb gauge}\label{sec:casimirCoulomb1}
\subsubsection{Ghost contributions}
Let us start with the ghost contributions coming from
\begin{equation*}
	S_{c,\bar{c},\eta,\bar{\eta}} = \int \dd[4]{x} \bigg[ \bar{c} \vec{\partial}^2 c + \left( \bar{\eta}^\rho c + \eta^\rho \bar{c}\right) \delta(z-z^\rho) \bigg],
\end{equation*}
where we have used that \(\frac{\delta \mathcal{F}[A]}{\delta A_\mu} = \delta_{\mu i} \partial_i\) for Coulomb gauge. The path integral in \((c,\bar{c})\) is Gaussian, so they can readily be integrated out using \eqref{eq:gaussian-grassmann}. Since we are only interested in contributions to \(\mathcal{E}_{\text{Cas}}\) that depend on the inter-plate distance \(L\), we can drop the factor \(\det \vec{\partial}^2\) (which simply yields an infinite constant anyhow):
\begin{equation*}
	Z_{c,\bar{c},\eta,\bar{\eta}} = C \exp \left( - \int \dd[4]{x} \bar\eta^\rho(\cev{x}) \delta(z-z^\rho) \frac{1}{\vec{\partial}^2} \eta^\sigma(\cev{x}) \delta(z-z^\sigma) \right),
\end{equation*}
and the infinite constant \(C\) will be simply omitted from now on. We can evaluate this integral by going to Fourier space (see \eqref{eq:fourier-convention} for our conventions). The integral then becomes (see (\ref{ap:ghost-action-Coulomb},~\ref{ap:beta-prop-coulomb}))
\begin{equation}\label{eq:eta-prop-coulomb}
	S_{\eta,\bar\eta} =- \int \frac{\dd[4]{k}}{(2\pi)^4}\bar\eta^\rho(\cev{k})e^{i k_z z^\rho} \frac{1}{\vec{k}^2} \eta^\sigma(-\cev{k})e^{i k_z z^\sigma}=  -\int \frac{\dd[3]{\cev{k}}}{(2\pi)^3}\bar\eta^\rho(\cev{k}) \mathbb{H}^{\rho\sigma} \eta^\sigma (-\cev{k}),
\end{equation}
where in the last equality one can perform the \(k_z\)-integral using the standard integrals \eqref{eq:kz-integrals}. 
Consequently
\begin{equation}\label{eq:ghost-contributions}
	Z_{c,\bar{c},\eta,\bar{\eta}} = \det \mathbb{H}=\exp \left( \int \frac{\dd[3]{\cev{k}}}{(2\pi)^3} \ln | \mathbb{H}_k |  \right)=1,
\end{equation}
see (\ref{ap:det-rule},~\ref{ap:ghost1}).
We thus find that the ghosts do not contribute to the Casimir energy.

\subsubsection{Non-ghost contributions}\label{sec:non-ghost-coulomb}
For the non-ghost contributions, we can start from the action \eqref{eq:action-h-out}. Indeed, the \(h\)-propagator did not contain any \(L\)-dependence, so we can omit its determinant: \(Z_{A,h,b,\gamma} = Z_{A,b,\gamma}\). If we inspect the term quadratic in \(\gamma\), we see that it is proportional to \(\delta(z-z^\rho) \delta(z-z^\sigma)\), which equals zero if \(\rho\neq\sigma\), and equals \(\delta(0)\) if \(\rho=\sigma\). 
Using \(\delta(0) = 0\) in gauge invariant dimensional regularization (see e.g.~\cite[Eq.~(4.2.6)]{Collins:1984xc} and \cite[below Eq.~(10.9)]{Zinn-Justin:2002ecy}), the term quadratic in \(\gamma\) vanishes identically. In Coulomb gauge, the action thus becomes
\begin{equation}\label{eq:non-ghost-action}
	S_{A,b,\gamma} = \int \dd[4]{x} \bigg[ \frac{1}{4} F_{\mu\nu} F_{\mu\nu} + \frac{1}{2\xi} \left( \partial_i A_i \right)^2 + \frac{1}{\xi} \left( \partial_i A_i \right) \gamma^\rho(\cev{x}) \delta(z-z^\rho) + \Big( b_t^\rho(\cev{x}) A_t + b_a^\rho(\cev{x}) F_{az} \Big) \delta(z-z^\rho) \bigg].
\end{equation}
After partial integration and rearranging some terms, this action can be brought in Gaussian form in Fourier space:
\begin{equation}\label{eq:action-A}
S_{A,b,\gamma} =	\int \frac{\dd[4]{k}}{(2\pi)^4} \bigg[\frac12 A_\mu(k) \mathbb{K}_{\mu\nu} A_\nu(-k) + A_\mu(k) v_\mu(-k) \bigg],
\end{equation}
for some quadratic operator \(\mathbb{K}\) and the vector \(v\) a linear combination of \(b\) and \(\gamma\) (see (\ref{ap:non-ghost-action}-\ref{ap:def-v})). One can then integrate out the \(A\)-fields using \eqref{eq:gaussian-scalar}. The determinant factor does not depend on \(L\), so we can omit it: \(Z_{A,b,\gamma} = Z_{b,\gamma}\). The left-over action now reads
\begin{equation*}
	S_{b,\gamma} = -\frac12 \int \frac{\dd[4]{k}}{(2\pi)^4} \bigg[ v_\mu(k) (\mathbb{K}^{-1})_{\mu\nu} v_\nu(-k) \bigg].
\end{equation*}
Since \(v\) contains linear terms both in \(b\) and in \(\gamma\), the action will consist of a term quadratic in \(b\), a term quadratic in \(\gamma\) (which turns out to vanish), and a mixing term (see (\ref{ap:action-b-gamma-1}-\ref{ap:action-b-gamma})):
%@
\begin{equation}\label{eq:action-b-gamma}
	 S_{b,\gamma} = -\frac12 \int \frac{\dd[3]{\cev{k}}}{(2\pi)^3} \Big( b_t^\rho(\cev{k}) \mathbb{M}^{\rho\sigma} b_t^\sigma(-\cev{k}) +  b_a^\rho(\cev{k}) \mathbb{N}^{\rho\sigma}_{ab} b_b^\sigma(-\cev{k}) + b_t^\rho(\cev{k}) \mathbb{O}^{\rho\sigma} \gamma^\rho(-\cev{k}) \Big).
\end{equation}
Because \(b_a\) decouples from \(b_t\) and \(\gamma\), we can calculate both path integrals independently.
Thus (see (\ref{ap:Zba}-\ref{ap:ZAhbgamma}))
\begin{eqnarray}
Z_{b,\gamma}  &=& Z_{b_a} Z_{b_t,\gamma}=\frac{1}{\sqrt{\det \mathbb{N}}}
\left( \frac{1}{\sqrt{\det \mathbb{M}}} \int \mathcal{D} \gamma e^{-S_\gamma} \right) \nonumber\\
 &=&\frac{1}{\sqrt{\det \mathbb{N}}}
 \frac{1}{\sqrt{\det \mathbb{M}}} \frac{1}{\sqrt{\det \left( \mathbb{O} \cdot \mathbb{M}^{-1} \cdot \mathbb{O} \right)}} = 
 \frac{1}{\sqrt{\det \mathbb{N}}}\frac{1}{\det \mathbb{O}},\nonumber
\end{eqnarray}
where
$ S_\gamma = -\frac12  \int \frac{\dd[3]{\cev{k}}}{(2\pi)^3}
	\gamma^\tau(\cev{k})   \mathbb{O}^{\tau\rho}
	(\mathbb{M}^{-1})^{\rho\sigma}
	\mathbb{O}^{\sigma\lambda} \gamma^\lambda(-\cev{k}).$
Computing the determinants, we get:
\begin{equation*}
	Z_{A,h,b,\gamma} = \frac{1}{\sqrt{\det \mathbb{N}}} \frac{1}{\det \mathbb{O}}=
	 \exp \left[ \frac{\pi^2}{720 L^3} \right]\cdot 1.
\end{equation*}
%@@@
 %@@@
We thus find the Casimir energy density per unit area to be
\begin{equation*}
	\mathcal{E}_{\text{Cas}} = - \frac{\pi^2}{720 L^3},
\end{equation*}
and, taking the derivative with respect to \(L\), the Casimir force per unit area
\begin{equation*}
	\mathcal{F}_{\text{Cas}} = - \frac{\pi^2}{240 L^4}.
\end{equation*}
This is the usual attractive Casimir force for two PEC plates \cite{Bordag:1983zk} (or two PMC plates \cite{Dudal:2020yah}).

Let us give a couple of comments regarding this result. To start, it is no surprise that the Casimir energy exhibits an \(L^{-3}\) behavior. Indeed, an energy density per unit area has mass dimension \(+3\), and \(L\) is the only dimensionful parameter in the theory. As such, the only question was what the prefactor would be.

We can also try to give some physical intuition regarding the fact that we find the exact same prefactor as for two PEC plates. Expression \eqref{ap:2polarizations} can be interpreted as two degrees of freedom (DOF) contributing to the Casimir energy (because of the square inside the logarithm). Here, these DOFs are \(b_x\) and \(b_y\). Looking at \eqref{ap:Z-bt_gamma}, one could say that the third ``edge mode'' DOF \(b_t\) gets annihilated by the ``conjugate edge mode'' DOF \(\gamma\). Let us compare this situation to the PMC case \cite{Dudal:2020yah} where the mechanism at work was slightly different. There, all three DOFs \(b_t,b_x,b_y\) were coupled in the boundary action, but this effective boundary theory had a local gauge redundancy under \(\cev{b} \rightarrow \cev{b} + a \cev{k}\). Fixing this emerging gauge freedom removed one DOF, such that only two DOFs could contribute to the Casimir energy, resulting in the same outcome as for the DEM plates.

Next, let us discuss the issues that arise when one does not introduce the BRST fields, but tries to calculate the Casimir energy directly from the action \eqref{eq:action-no-ghosts} which is not gauge or BRST  invariant. One then follows the exact same calculations as in this section, but since there is no \(\gamma\), the last mixing term in the action \eqref{eq:action-b-gamma} is not there. The \(b_a\) contribution is identical as for the full BRST invariant action, but the \(b_t\) contribution gives trouble. Indeed, this contribution is \(1/\sqrt{\det \mathbb{M}}\), an expression that contains the gauge parameter \(\xi\). It is very satisfying to see that the new field \(\gamma\), that was introduced to restore BRST invariance, neatly cancels this contribution, thereby removing the gauge parameter from the Casimir energy.

Lastly, remember that in section \ref{sec:edgeDynamics} we posited the form of \(S_{b,\gamma}\) for one plate in \eqref{eq:2D-lapl}, i.e.~we claimed that the \(b_t\)-propagator is of the form \(\frac{1}{\sqrt{-\nabla^2_{2D}}}\), and that \(\frac{\partial_t}{\sqrt{-\nabla^2_{2D}}} \gamma\) sources \(b_t\). We can now prove this statement by inspecting the action \eqref{eq:action-b-gamma} for a one-plate configuration with \(\xi =0\). By performing the \(k_z\)-integral in \eqref{ap:action-b-gamma-3}, we then get \(\mathbb{M} = \frac{1}{2\sqrt{k_x^2+k_y^2}}\), which is indeed (proportional to) the Fourier transform of \(\frac{1}{\sqrt{-\nabla^2_{2D}}}\). For one plate, the mixing operator \eqref{ap:def-O} becomes \(\mathbb{O} = \frac{-i k_t}{2\sqrt{k_x^2 + k_y^2}} \), which is (proportional to) the Fourier transform of \(\frac{\partial_t}{\sqrt{-\nabla^2_{2D}}} \).

\subsection{Casimir energy in linear covariant gauge}\label{sec:casimirLandau1}
Let us calculate the Casimir energy a second time, but now in linear covariant gauge \(\mathcal{F}[A] = \partial_\mu A_\mu\). The calculation is completely analogous to the Coulomb case, so we will mostly focus on the differences we encounter.

\subsubsection{Ghost contributions}
We can copy the argument for the Coulomb case, simply substituting \(\vec{\partial}^2 \rightarrow \partial^2\), or equivalently \(\vec{k}^2 \rightarrow k^2\). After integrating out \((c,\bar{c})\), the \((\eta,\bar\eta)\)-action in Fourier space \eqref{eq:eta-prop-coulomb} thus becomes
\begin{equation*}
	S_{\eta,\bar\eta} = - \int \frac{\dd[4]{k}}{(2\pi)^4}\bar\eta^\rho(\cev{k})e^{i k_z z^\rho} \frac{1}{k^2} \eta^\sigma(-\cev{k})e^{i k_z z^\sigma}.
\end{equation*}
Applying the \(k_z\)-integral \eqref{eq:kz-integrals}, this becomes
\begin{equation*}
	S_{\eta,\bar\eta} = -\int \frac{\dd[3]{\cev{k}}}{(2\pi)^3}\bar\eta^\rho(\cev{k}) \frac{e^{|\cev{k}| \left| z^\rho - z^\sigma \right|}}{2|\cev{k}|} \eta^\sigma(-\cev{k}).
\end{equation*}
It is important to note that, this time, the integrand  does depend on \(k_t\),  so we cannot conclude that \(Z_{c,\bar{c},\eta,\bar\eta} = 1\). Instead, we find
\begin{equation*}
	Z_{c,\bar{c},\eta,\bar\eta} = \exp \left[ \int \frac{\dd[3]{\cev{k}}}{(2\pi)^3} \log \left( 1 - e^{-2|\cev{k}|L} \right) \right].
\end{equation*}
Comparing this to \eqref{ap:2polarizations}, we see that the ghost contribution will exactly cancel the contribution of two DOFs. Since we have to end up with the same result as in Coulomb gauge, i.e. two contributing DOFs, we will need to find two extra DOFs that contribute.

\subsubsection{Non-ghost contributions}\label{sec:non-ghost-landau}
For the non-ghost contributions, we can modify the calculations done in \ref{sec:non-ghost-coulomb} by substituting \(\partial_i A_i \rightarrow \partial_\mu A_\mu\). Doing so, we again find a Gaussian path integral \eqref{eq:action-A} for \(A\), but with
\begin{equation*}
	\mathbb{K}_{\mu\nu} = \left(k^2 \delta_{\mu\nu} -\left(1-\frac{1}{\xi} \right) k_\mu k_\nu \right)
\end{equation*}
and
\begin{equation*}
	\begin{cases}
		v_t(-k) &= \left( b_t^\rho(-\cev{k}) - \frac{i}{\xi} k_t \gamma^\rho(-\cev{k}) \right) e^{i k_z z^\rho} \\
		v_a(-k) &= \left( i k_z b_a^\rho(-\cev{k}) - \frac{i}{\xi} k_a \gamma^\rho(-\cev{k}) \right) e^{i k_z z^\rho} \\
		v_z(-k) &= \left(-i k_a b_a^\rho(-\cev{k}) - \frac{i}{\xi} k_z \gamma^\rho(-\cev{k}) \right) e^{i k_z z^\rho}
	\end{cases}.
\end{equation*}
Using that
\begin{equation*}
	(\mathbb{K}^{-1})_{\mu\nu} = \frac{\delta_{\mu\nu}}{k^2} + \frac{\xi-1}{k^4} k_\mu k_\nu,
\end{equation*}
the path integral for \(A\) can be done, and in the resulting action, the quadratic term in \(\gamma\) vanishes once again. One can straightforwardly check that the action is again of the form
\begin{equation*}
	S_{b,\gamma} = -\frac12 \int \frac{\dd[3]{\cev{k}}}{(2\pi)^3} \Big( b_t^\rho(\cev{k}) \mathbb{M}^{\rho\sigma} b_t^\sigma(-\cev{k}) +  b_a^\rho(\cev{k}) \mathbb{N}^{\rho\sigma}_{ab} b_b^\sigma(-\cev{k}) + b_t^\rho(\cev{k}) \mathbb{O}^{\rho\sigma} \gamma^\sigma(-\cev{k}) \Big).
\end{equation*}
The \(b_t\) propagator \(\mathbb{M}\) is different than in Coulomb gauge, but here as well, its contribution will nicely drop out. The \(b_a\) propagator \(\mathbb{N}\) is the exact same \eqref{ap:def-N} as in Coulomb gauge. The mixing operator \(\mathbb{O}\) has changed by the substitution \(\sqrt{k_x^2+k_y^2} \rightarrow |\cev{k}|\), resulting in \(\mathbb{O}^{\rho\sigma} = \frac{-i k_t}{2|\cev{k}|}  e^{ - |\cev{k}| |z^\rho - z^\sigma|} \).

We still have
\begin{equation*}
	Z_{b,\gamma} = \frac{1}{\sqrt{\det \mathbb{N}}}
 \frac{1}{\sqrt{\det \mathbb{M}}} \frac{1}{\sqrt{\det \left( \mathbb{O} \cdot \mathbb{M}^{-1} \cdot \mathbb{O} \right)}} = \frac{1}{\sqrt{\det \mathbb{N}}}\frac{1}{\det \mathbb{O}},
\end{equation*}
but the second determinant is not trivial any more:
\begin{equation*}
	\frac{1}{\det \mathbb{O}} = \exp \left[ -\int \frac{\dd[3]{\cev{k}}}{(2\pi)^3} \log \left( 1 - e^{-2 |\cev{k}|L} \right) \right].
\end{equation*}
This exactly gives us the two extra DOF contributions we were looking for. In conclusion: we find that the contributions from the ghost DOFs and the \((b_t,\gamma)\) DOFs exactly cancel each other, leaving the same two DOFs from \(b_a\). The Casimir energy density we find is thus again the usual \(\mathcal{E}_{\text{Cas}} = - \frac{\pi^2}{720 L^3}\).

\section{Conclusion and outlook}\label{ch:conclusion}

We have derived the Casimir energy for two infinite parallel plates carrying DEM conditions, and have found the exact same value as for two perfectly conducting plates. Depending on what gauge function is chosen, different fields contribute to the energy. Our method has been to introduce the boundary conditions into the action using Lagrange multipliers, and then using functional methods to obtain the Casimir energy. In order to find a gauge invariant result, we needed to introduce boundary ghosts and a boundary scalar to restore BRST invariance on the boundary. We have explicitly discussed the correspondence of this BRST invariant theory with the original DEM formulation.

It is worth noting the benefit that several non-trivial observations follow automatically from the requirement of BRST invariance. A first example is the fact that the physical part of the boundary electrical flux is non-dynamical (cfr.~\eqref{eq:Gaussdt}), which follows directly from the BRST  transformation \(s\bar\eta = -\partial_t b_t\). As noted in the main text already, this agrees with the discussion of \cite{Riello:2021lfl}, see also \cite{Gomes:2019xto}, relating edge modes to (non-dynamical) electric fluxes.

A second example is the fact that one automatically is led to introduce the conjugate edge mode \(\gamma\), for which there otherwise would not be any obvious direct raison d'\^etre from a Lagrangian perspective. Interestingly enough, in our BRST formulation, this conjugate mode is part of a BRST doublet and thus apparently a physically trivial degree of freedom.

The study of the Casimir effect for DEM boundary conditions has led us to several interesting open questions. A first one is to investigate the physical degrees of freedom contained in the BRST  invariant theory by constructing the canonical Fock space and projecting out the BRST trivial part. At first sight, the physical subspace seems to consist of the usual two transverse gauge polarizations, supplemented with the extra degree of freedom encoded in the $b_t$-sector, at least if $\partial_t b_t= 0$. Related is the question whether or not the precise interpretation of edge modes changes in gauges different than Coulomb.

In addition, it would be interesting to further generalize the boundary Lagrange multiplier field method---again in conjunction with the BRST symmetry---to generic manifolds with boundaries, which would allow a more in-depth study of edge modes in more general systems.

Another natural extension of the current work is to consider the Casimir energy for DEM conditions in combination with other sets of boundary conditions. In particular, two-plate configurations with DEM conditions on one plate and more ``standard'' conditions (such as PEC or PMC) on the other plate come to mind. Additionally, one might also look for other combinations of  DEM and/or PEC-/PMC-like boundary conditions, inspired by the boundary variations of the classical action, including more general PEMC boundary conditions as studied in \cite{lindell2005perfectelectromagneticconductor,Rode:2017yqy,Dudal:2024PEMC} or with a chiral medium between the plates \cite{Fukushima:2019sjn,Canfora:2022xcx,Oosthuyse:2023mbs}. 

\section*{Acknowledgments}
We thank A.~Ball for suggesting to study the Casimir energy in relation to DEM, and for some interesting discussions. The work of D.~Dudal and T.~Oosthuyse was supported by KU Leuven IF project C14/21/087. The work of S.~Stouten was funded by FWO PhD-fellowship fundamental research (file number: 1132823N). L.~Rosa acknowledges the Ministero dell’Universit\`a e della
Ricerca (MUR), PRIN2022 program (Grant PANTHEON 2022E2J4RK) for partial support. F.~Canfora has been funded by FONDECYT Grant No.~1240048.

\appendix
\section{Fourier conventions, \(k_z\)-integrals, Gaussian integrals}\label{ch:appendix}
We will follow the Fourier convention used in \cite{Peskin:1995ev}
\begin{equation}\label{eq:fourier-convention}
	X_i(x) = \int \frac{\dd[d]{k}}{(2\pi)^d} \hat{X}_i(k)e^{-ik\cdot x}
\end{equation}
for a \(d\)-dimensional vector field \(X\), but we will drop the hat in order to not unnecessarily overload notations. In this convention, the Dirac delta is Fourier transformed as
\begin{equation*}
	\delta(z-z^\rho) = \int \frac{\dd{k_z}}{2\pi} e^{-ik_z(z-z^\rho)}.
\end{equation*}

Let us list the \(k_z\)-integrals we encounter in our calculations
\begin{align}
	\begin{split}\label{eq:kz-integrals}
		&\int \frac{\dd{k_z}}{2\pi} \frac{e^{i k_z(z^\rho - z^\sigma)} }{k^2} = \frac{1}{2|\cev{k}|}e^{-|\cev{k}| |z^\rho - z^\sigma|}, \\
		%&\int \frac{\dd{k_z}}{2\pi} \frac{k_z e^{i k_z(z^\rho - z^\sigma)} }{k^2} =
		%-\frac{i}{2} \varepsilon^{\rho\sigma} e^{-|\cev{k}| |z^\rho - z^\sigma|}, \\
		&\int \frac{\dd{k_z}}{2\pi} \frac{k_z^2 e^{i k_z(z^\rho - z^\sigma)} }{k^2} = -\frac{|\cev{k}|}{2}e^{-|\cev{k}| |z^\rho - z^\sigma|}, \\
		&\int \frac{\dd{k_z}}{2\pi} \frac{e^{i k_z(z^\rho - z^\sigma)} }{k^4} = \frac{1+|\cev{k}| |z^\rho - z^\sigma|}{4|\cev{k}|^3}  e^{-|\cev{k}| |z^\rho - z^\sigma|}, \\
		%&\int \frac{\dd{k_z}}{2\pi} \frac{k_z e^{i k_z(z^\rho - z^\sigma)} }{k^4} =
		%-\frac{i}{4} \varepsilon^{\rho\sigma} \frac{|z^\rho - z^\sigma|}{|\cev{k}|} e^{-|\cev{k}| |z^\rho - z^\sigma|},\\
		&\int \frac{\dd{k_z}}{2\pi} \frac{k_z^2 e^{i k_z(z^\rho - z^\sigma)} }{k^4} = \frac{1-|\cev{k}| |z^\rho - z^\sigma|}{4|\cev{k}|}  e^{-|\cev{k}| |z^\rho - z^\sigma|},
	\end{split}
\end{align}
%where we set \(\varepsilon^{-+} = 1\), and no summation over \(\rho\) or \(\sigma\) is implied in the right-hand side.
where no summation over \(\rho\) or \(\sigma\) is implied in the right-hand side.

The Gaussian path integral for scalar variables \(\Phi\) is given by
\begin{equation}\label{eq:gaussian-scalar}
	\int \mathcal{D}\Phi \exp \left[\int \dd{x} \Phi_i K_{ij} \Phi_j + \Phi_i J_i \right]
	= C \frac{1}{\sqrt{\det K}} \exp \left( - \int \dd{x} J_i (K^{-1})_{ij} J_j \right),
\end{equation}
where \(C\) is an infinite constant which can always be omitted in practice.

The Gaussian path integral for Grassmann variables \(\Theta,\bar\Theta\) is given by
\begin{equation}\label{eq:gaussian-grassmann}
	\int \mathcal{D}\Theta \mathcal{D}\bar{\Theta} \exp \left[ \int \dd{x} \bar\Theta_i K_{ij} \Theta_j + \bar\Theta_i \eta_i + \Theta_i \bar\eta_i \right]
	= C \det K \exp \left( -\int \dd{x} \bar\eta_i (K^{-1})_{ij} \eta_j \right).
\end{equation}
\section{Detailed computation of the Casimir energy in Coulomb gauge}\label{ch2:appendix}

\subsubsection{Ghost contributions}
We depart from the action \eqref{eq:eta-prop-coulomb} for \((\eta,\bar\eta)\)
\begin{equation}\label{ap:ghost-action-Coulomb}
	S_{\eta,\bar\eta} = - \int \frac{\dd[4]{k}}{(2\pi)^4}\bar\eta^\rho(\cev{k})e^{i k_z z^\rho} \frac{1}{\vec{k}^2} \eta^\sigma(-\cev{k})e^{i k_z z^\sigma},
\end{equation}
in which we can do the \(k_z\)-integral using the standard integrals \eqref{eq:kz-integrals}. This yields
\begin{equation}\label{ap:beta-prop-coulomb}
	S_{\eta,\bar\eta} = -\int \frac{\dd[3]{\cev{k}}}{(2\pi)^3}\bar\eta^\rho(\cev{k}) \frac{e^{\sqrt{k_x^2+k_y^2} \left| z^\rho - z^\sigma \right|}}{2\sqrt{k_x^2+k_y^2}} \eta^\sigma(-\cev{k})=:  -\int \frac{\dd[3]{\cev{k}}}{(2\pi)^3}\bar\eta^\rho(\cev{k}) \mathbb{H}^{\rho\sigma} \eta^\sigma (-\cev{k}).
\end{equation}
Thus, using \eqref{eq:gaussian-grassmann}, we find (see for example  \cite[Eq.~(11.71)]{Peskin:1995ev})
\begin{equation}\label{ap:det-rule}
	Z_{c,\bar{c},\eta,\bar{\eta}} = \det \mathbb{H}=\exp \left( \int \frac{\dd[3]{\cev{k}}}{(2\pi)^3} \ln | \mathbb{H}_k |  \right),
\end{equation}
where \(| \mathbb{H}_k |\) denotes the matrix determinant of the operator \(\mathbb{H}\) in momentum representation. Inspecting the momentum representation \(\mathbb{H}_k\) in \eqref{ap:beta-prop-coulomb}, we see that it does not depend on \(k_t\). 
Consequently, the ghost path integral becomes
\begin{equation}\label{ap:ghost1}
	Z_{c,\bar{c},\eta,\bar{\eta}} = \exp \left( \int \frac{\dd{k_t}}{2\pi} \cdot \int \frac{\dd[2]{k}}{(2\pi)^2} \ln \left| \mathbb{H}_k \right|  \right)=1.
\end{equation}
Indeed, the factor \(\int \frac{\dd{k_t}}{2\pi}\) yields an infinite \(\delta(0)\), which, however, is equal to zero in gauge invariant dimensional regularization (see e.g.~\cite[Eq.~(4.2.6)]{Collins:1984xc} and \cite[below Eq.~(10.9)]{Zinn-Justin:2002ecy}).

\subsubsection{Non-ghost contributions}\label{ap:non-ghost-coulomb}
Let us depart from the action \eqref{eq:non-ghost-action}
\begin{equation}\label{ap:non-ghost-action}
	S_{A,b,\gamma} = \int \dd[4]{x} \bigg[ \frac{1}{4} F_{\mu\nu} F_{\mu\nu} + \frac{1}{2\xi} \left( \partial_i A_i \right)^2 + \frac{1}{\xi} \left( \partial_i A_i \right) \gamma^\rho(\cev{x}) \delta(z-z^\rho) + \Big( b_t^\rho(\cev{x}) A_t + b_a^\rho(\cev{x}) F_{az} \Big) \delta(z-z^\rho) \bigg].
\end{equation}
Gathering the terms quadratic in \(A\) and doing partial integration where necessary, we find
\begin{align*}
	\begin{split}
		S_{A,b,\gamma} &= \int \dd[4]{x} \bigg[ -\frac12 A_\mu \left( \delta_{\mu\nu} \partial^2 - \partial_\mu \partial_\nu + \frac{1}{\xi} \delta_{\mu i}\delta_{\nu j} \partial_i \partial_j \right) A_\nu \\
		&\qquad\qquad\qquad+ \frac{1}{\xi} \left( \partial_i A_i \right) \gamma^\rho(\cev{x}) \delta(z-z^\rho) + \Big( b_t^\rho(\cev{x}) A_t + b_a^\rho(\cev{x}) F_{az} \Big) \delta(z-z^\rho) \bigg].		
	\end{split}
\end{align*}
As a next step, we want to integrate out \(A\). This is most easily done in Fourier space, where the action becomes
\begin{align*}
	\begin{split}
		S_{A,b,\gamma} &= \int \frac{\dd[4]{k}}{(2\pi)^4} \bigg[ \frac12 A_\mu(k) \left( \delta_{\mu\nu} k^2 - k_\mu k_\nu + \frac{1}{\xi} \delta_{\mu i}\delta_{\nu j} k_i k_j \right) A_\nu(-k) \\
		&\qquad\qquad + \left( -\frac{i}{\xi} \left( k_i A_i(k) \right) \gamma^\rho(-\cev{k}) +  A_t(k) b_t^\rho(-\cev{k}) - i \left( k_a A_z(k) - k_z A_a(k) \right) b_a^\rho(-\cev{k}) \right) e^{i k_z z^\rho} \bigg].
	\end{split}
\end{align*}
This action is of the Gaussian form
\begin{equation*}
	\int \frac{\dd[4]{k}}{(2\pi)^4} \bigg[\frac12 A_\mu(k) \mathbb{K}_{\mu\nu} A_\nu(-k) + A_\mu(k) v_\mu(-k) \bigg],
\end{equation*}
with
\begin{equation}
	\mathbb{K}_{\mu\nu} = \left(k^2 \delta_{\mu\nu} - k_\mu k_\nu + \frac{1}{\xi} k_i \delta_{i\mu} k_j \delta_{j\nu} \right)
\end{equation}
and
\begin{equation}\label{ap:def-v}
	\begin{cases}
		v_t(-k) &= b_t^\rho(-\cev{k}) e^{i k_z z^\rho} \\
		v_a(-k) &= \left( i k_z b_a^\rho(-\cev{k}) - \frac{i}{\xi} k_a \gamma^\rho(-\cev{k}) \right) e^{i k_z z^\rho} \\
		v_z(-k) &= \left(-i k_a b_a^\rho(-\cev{k}) - \frac{i}{\xi} k_z \gamma^\rho(-\cev{k}) \right) e^{i k_z z^\rho}
	\end{cases}.
\end{equation}
We can now integrate out the \(A\)-fields using \eqref{eq:gaussian-scalar}. The determinant factor does not depend on \(L\), so we can omit it: \(Z_{A,b,\gamma} = Z_{b,\gamma}\). The left-over action now reads
\begin{equation*}
	S_{b,\gamma} = -\frac12 \int \frac{\dd[4]{k}}{(2\pi)^4} \bigg[ v_\mu(k) (\mathbb{K}^{-1})_{\mu\nu} v_\nu(-k) \bigg],
\end{equation*}
with
\begin{equation*}
	(\mathbb{K}^{-1})_{\mu\nu} = \frac{\delta_{\mu\nu}}{k^2} + \frac{1}{k^2 \vec{k}^2} \left( \xi \frac{k^2}{\vec{k}^2} k_\mu k_\nu  + k_t^2 \delta_{t \mu} \delta_{t \nu} - k_i \delta_{i \mu} k _j \delta_{j \nu}   \right).
\end{equation*}
Since \(v_\mu\) contains terms both in \(b\) and in \(\gamma\), the action will consist of a term quadratic in \(b\), a term quadratic in \(\gamma\), and a mixing term.

Firstly, let us inspect the term quadratic in \(\gamma\). One has
\begin{align}
	\begin{split}\label{ap:action-b-gamma-1}
		S_\gamma^\text{quad} &= -\frac12 \int \frac{\dd[4]{k}}{(2\pi)^4} \frac{i}{\xi} k_i \gamma^\rho(\cev{k}) (\mathbb{K}^{-1})_{ij} \frac{-i}{\xi} k_j \gamma^\sigma(-\cev{k}) e^{i k_z (z^\rho - z^\sigma)} \\
		&= -\frac{1}{2\xi} \int \frac{\dd[4]{k}}{(2\pi)^4} \gamma^\rho(\cev{k}) \gamma^\sigma(-\cev{k}) e^{i k_z (z^\rho - z^\sigma)} \\
		&= -\frac{1}{2\xi} \int \frac{\dd[3]{\cev{k}}}{(2\pi)^3} \gamma^\rho(\cev{k}) \gamma^\sigma(-\cev{k}) \delta(z^\rho - z^\sigma),
	\end{split}
\end{align}
which is zero by the same argument as in the beginning of section \ref{sec:non-ghost-coulomb}.

Secondly, let us inspect the mixing term, which we can interpret as a source term for \(b\). After a bit of algebra, one finds that \(\gamma\) only couples to \(b_t\):
\begin{align}
	\begin{split}\label{ap:action-b-gamma-2}
		S_{b,\gamma}^\text{bilin} &= -\frac12 \int \frac{\dd[4]{k}}{(2\pi)^4} \frac{-i k_t}{\vec{k}^2} b_t^\rho(\cev{k}) \gamma^\sigma(-\cev{k}) e^{i k_z (z^\rho - z^\sigma)} \\
		&= -\frac12 \int \frac{\dd[3]{\cev{k}}}{(2\pi)^3} \frac{-i k_t}{2\sqrt{k_x^2 + k_y^2}}  e^{ - \sqrt{k_x^2 + k_y^2} |z^\rho - z^\sigma|} b_t^\rho(\cev{k}) \gamma^\sigma(-\cev{k}).
	\end{split}
\end{align}

Lastly, let us inspect the term quadratic in \(b\). After similar algebraic manipulations, one discovers that the \(b_t\) and \(b_a\) sectors decouple. Indeed, the quadratic term in \(b\) can be written as
\begin{align}
	\begin{split}\label{ap:action-b-gamma-3}
		S_{b}^\text{quad} &= -\frac12 \int \frac{\dd[4]{k}}{(2\pi)^4} \left( b_t^\rho(\cev{k}) \frac{\vec{k}^2 + \xi k_t^2}{\vec{k}^4} b_t^\sigma(-\cev{k}) +  b_a^\rho(\cev{k}) \frac{k_a k_b + k_z^2 \delta_{ab}}{k^2} b_b^\sigma(-\cev{k}) \right) e^{i k_z (z^\rho - z^\sigma)} \\
		&= -\frac12 \int \frac{\dd[3]{\cev{k}}}{(2\pi)^3} \Big( b_t^\rho(\cev{k}) \mathbb{M}^{\rho\sigma} b_t^\sigma(-\cev{k}) +  b_a^\rho(\cev{k}) \mathbb{N}^{\rho\sigma}_{ab} b_b^\sigma(-\cev{k}) \Big).	
	\end{split}
\end{align}
To find explicit expressions for \(\mathbb{M}\) and \(\mathbb{N}\), one needs to perform the \(k_z\)-integral using \eqref{eq:kz-integrals}. We will not be writing down the expression for the \(b_t\)-propagator \(\mathbb{M}\) since we will not need it. We do, however, need the \(b_a\)-propagator \(\mathbb{N}\), which is given by
\begin{equation}\label{ap:def-N}
	\mathbb{N}^{\rho\sigma}_{ab} = \frac{k_a k_b - \cev{k}^2 \delta_{ab}}{2|\cev{k}|} e^{-|\cev{k}| |z^\rho-z^\sigma|}.
\end{equation}

We can now gather the terms (\ref{ap:action-b-gamma-1},~\ref{ap:action-b-gamma-2},~\ref{ap:action-b-gamma-3}) to obtain
\begin{equation}\label{ap:action-b-gamma}
	 S_{b,\gamma} = -\frac12 \int \frac{\dd[3]{\cev{k}}}{(2\pi)^3} \Big( b_t^\rho(\cev{k}) \mathbb{M}^{\rho\sigma} b_t^\sigma(-\cev{k}) +  b_a^\rho(\cev{k}) \mathbb{N}^{\rho\sigma}_{ab} b_b^\sigma(-\cev{k}) + b_t^\rho(\cev{k}) j_t^\rho(-\cev{k}),  \Big)
\end{equation}
where we introduced the source \(j_t^\rho(-\cev{k}) = \frac{-i k_t}{2\sqrt{k_x^2 + k_y^2}}  e^{ - \sqrt{k_x^2 + k_y^2} |z^\rho - z^\sigma|} \gamma^\sigma(-\cev{k})\).

Since \(b_a\) decouples from \(b_t\) and \(\gamma\), we can calculate both path integrals independently: \(Z_{b,\gamma} = Z_{b_a} Z_{b_t,\gamma}\). The first one can immediately be found using \eqref{eq:gaussian-scalar}:
\begin{equation}\label{ap:Zba}
	Z_{b_a} = \frac{1}{\sqrt{\det \mathbb{N}}}.
\end{equation}
The second one is also a Gaussian integral, but with a source, yielding
\begin{equation*}
	Z_{b_t,\gamma} = \frac{1}{\sqrt{\det \mathbb{M}}} \int \mathcal{D} \gamma e^{-S_\gamma},
\end{equation*}
with
\begin{align*}
	S_\gamma &= \frac12 \int \frac{\dd[3]{\cev{k}}}{(2\pi)^3} j_t^\rho(\cev{k}) (\mathbb{M}^{-1})^{\rho\sigma} j_t^\sigma(-\cev{k}) \\
	&= -\frac12  \int \frac{\dd[3]{\cev{k}}}{(2\pi)^3}
	\gamma^\tau(\cev{k})   \mathbb{O}^{\tau\rho}
	(\mathbb{M}^{-1})^{\rho\sigma}
	\mathbb{O}^{\sigma\lambda} \gamma^\lambda(-\cev{k}),
\end{align*}
where we have introduced
\begin{equation}\label{ap:def-O}
	\mathbb{O}^{\tau\rho} =  \frac{i k_t}{2\sqrt{k_x^2 + k_y^2}}  e^{ - \sqrt{k_x^2 + k_y^2} |z^\tau - z^\rho|}.
\end{equation}
We can now perform the final Gaussian path integral over \(\gamma\), yielding
\begin{equation}\label{ap:Z-bt_gamma}
	Z_{b_t,\gamma} = \frac{1}{\sqrt{\det \mathbb{M}}} \frac{1}{\sqrt{\det \left( \mathbb{O} \cdot \mathbb{M}^{-1} \cdot \mathbb{O} \right)}} = \frac{1}{\det \mathbb{O}}.
\end{equation}
Note that \(\mathbb{M}\) indeed drops out automatically. Gathering all non-ghost contributions, we have
\begin{equation}\label{ap:ZAhbgamma}
	Z_{A,h,b,\gamma} = \frac{1}{\sqrt{\det \mathbb{N}}} \frac{1}{\det \mathbb{O}}
\end{equation}

Since we have explicit expressions for \(\mathbb{N}\) and \(\mathbb{O}\) in momentum space (resp.~\eqref{ap:def-N} and \eqref{ap:def-O}), calculating the determinant is straightforward using \eqref{ap:det-rule}:
\begin{align}
	\frac{1}{\sqrt{\det \mathbb{N}}} &= \exp \left[-\frac12 \int \frac{\dd[3]{\cev{k}}}{(2\pi)^3} \log | \mathbb{N}_k | \right]  \nonumber\\
	&= \exp \left[ -\frac12 \int \frac{\dd[3]{\cev{k}}}{(2\pi)^3} \log \left( (1 - e^{-2|\cev{k}|L})^2 \right) \right] \label{ap:2polarizations}  \\
	&= \exp \left[ \frac{\pi^2}{720 L^3} \right] \label{ap:N-contribution}
\end{align}
where we have omitted all factors independent of the inter-plate distance \(L\). Analogously
\begin{align*}
	\frac{1}{\det \mathbb{O}} &= \exp \left[ -\int \frac{\dd[3]{\cev{k}}}{(2\pi)^3} \log | \mathbb{O}_k | \right] \\
	&= \exp \left[ -\int \frac{\dd[3]{\cev{k}}}{(2\pi)^3} \log \left( 1 - e^{-2\sqrt{k_x^2+k_y^2}L} \right) \right] \\
	&= 1,
\end{align*}
because of the same reasoning as for the ghost contributions \eqref{ap:ghost1}. We thus find only one non-trivial contribution \eqref{ap:N-contribution} to the path integral. 

\bibliography{bibliography}

\end{document}